 \documentstyle[12pt]{article}

\def\fun#1#2{\lower3.6pt\vbox{\baselineskip0pt\lineskip.9pt
        \ialign{$\mathsurround=0pt#1\hfill##\hfil$\crcr#2\crcr\sim\crcr}}}

\renewcommand\({\left(}
\renewcommand\){\right)}
\renewcommand\[{\left[}
\renewcommand\]{\right]}

\newcommand\eq[1]{Eq.~(\ref{#1})}
\newcommand\eqs[2]{Eqs.~(\ref{#1}) and (\ref{#2})}
\newcommand\eqss[3]{Eqs.~(\ref{#1}), (\ref{#2}) and (\ref{#3})}

\newcommand\ee{\end{equation}}
\newcommand\be{\begin{equation}}
\newcommand\eea{\end{eqnarray}}
\newcommand\bea{\begin{eqnarray}}

\newcommand\TeV{\,\mbox{TeV}}
\newcommand\GeV{\,\mbox{GeV}}
\newcommand\MeV{\,\mbox{MeV}}

\newcommand\eV{\,\mbox{eV}}

\newcommand\msun{M_\odot}
\newcommand\mpl{M_{\rm P}}

\newcommand\mpsis{|m_\chi^2|}

\newcommand{\lsim}{\mbox{\raisebox{-.9ex}{~$\stackrel{\mbox{$<$}}{\sim}$~}}}

\newcommand{\gsim}{\mbox{\raisebox{-.9ex}{~$\stackrel{\mbox{$>$}}{\sim}$~}}}

\newcommand\diff{\mbox d}

\def\dslash{\not{\hbox{\kern-2pt $\partial$}}}
\def\Dslash{\not{\hbox{\kern-4pt $D$}}}
\def\Oslash{\not{\hbox{\kern-4pt $O$}}}
\def\Qslash{\not{\hbox{\kern-4pt $Q$}}}
\def\pslash{\not{\hbox{\kern-2.3pt $p$}}}
\def\kslash{\not{\hbox{\kern-2.3pt $k$}}}
\def\qslash{\not{\hbox{\kern-2.3pt $q$}}}

 \newtoks\slashfraction
 \slashfraction={.13}
 \def\slash#1{\setbox0\hbox{$ #1 $}
 \setbox0\hbox to \the\slashfraction\wd0{\hss \box0}/\box0 }

\def\ee{\end{equation}}
\def\be{\begin{equation}}

\def\calp{{\cal P}}
\def\calr{{\cal R}}
\def\calpr{{\calp_\calr}}

\newcommand\sub[1]{_{\rm #1}}

\begin{document}


\begin{center}
{\Large\bf Models of inflation liberated by the curvaton hypothesis}

\bigskip

{\Large   Konstantinos Dimopoulos$^*$$^\dag$ and David H.~Lyth$^*$}

\bigskip

$^*${\it Physics Department, Lancaster University, Lancaster LA1 4YB,
UK}

$^\dag${\it Department of Physics, University of Oxford, 
Keble Road, Oxford OX1 3RH, UK}

\begin{abstract}
It is usually supposed that inflation is of the slow-roll variety, and that
the inflaton generates the primordial curvature perturbation. According
to the curvaton hypothesis, inflation need not be slow-roll, and if it is
the inflaton generates a negligible curvature perturbation. We find that
the construction of slow-roll inflation models becomes much easier under
this hypothesis. Also, thermal inflation followed by fast-roll becomes 
viable, with no slow-roll inflation at all.
\end{abstract}


\end{center}

\section{Introduction}
The primordial density perturbation, responsible for the origin of structure
in the Universe, is dominated by its  adiabatic component though significant
isocurvature components are not ruled out. The adiabatic component is 
determined by the curvature perturbation $\zeta$ of uniform-density
slices of spacetime, which has an almost flat spectrum. The normalization
of the spectrum at the  scales explored by the CMB anisotropy
 is given by \cite{wmapparam}
\be
\calp_\zeta^\frac12 \simeq  5\times 10^{-5} \label{CMBnorm}
\,.
\ee
The spectral index $n\equiv 1+\diff\ln\calp_\zeta/\diff\ln k$
is
 in the (1-$\sigma$) range \cite{wmapparam,wmapsdss}
\be
n=0.97\pm 0.03
\,.
\label{nbound}
\ee

Since it is  present on super-horizon scales, 
the  primordial curvature perturbation originates presumably during 
an era of inflation, at the beginning of which the whole observable
Universe is inside the horizon. 
The usual hypothesis (which we shall call the inflaton hypothesis)
is that the curvature 
perturbation comes from the vacuum fluctuation of the 
inflaton field, defined in this context as the one whose value determines
the end of inflation.
This makes it quite difficult to construct sensible models of 
 slow--roll inflation \cite{book,treview},
 and of course it rules out completely the possibility
that the curvature perturbation might originate during
thermal inflation \cite{bg,thermal1,thermal2,thermalrest,hs} 
which does not
have an inflaton.

According to the inflaton hypothesis, 
the curvature perturbation has already
reached its observed value at the end of inflation and does not change
thereafter. The simplest alternative is to suppose that the curvature
perturbation is negligible at the end of inflation, being generated
later from the perturbation of some `curvaton'  field different from
the inflaton \cite{Lyth:2001nq}. (This possibility was actually noticed
much earlier in two papers \cite{sylvia,lm}, but it was not pursued
at the time by the authors or by the community.)
This curvaton paradigm  has attracted a lot of attention
\cite{andrew,mt2,fy,luw,sloth,hmy,hofmann,mormur,ekm,mwu,postma,fl,gl,%
kostasquint,giov,lu,dllr1,ejkm,03dl2,dllr2,dlnr,pm,john2,kkt,ekm03,hktt,%
john3,bmr,bmr2,giovannini03,john4,mazumdar,ad,armen,cdl}
because it opens up new possibilities both for observation and for
model-building.\footnote
{According to the scenario developed in the above papers, the curvature
perturbation is generated by the oscillation of the curvaton field.
A different idea \cite{decay} is that the field causing the 
curvature perturbation does so because its value determines the epoch
of reheating, and another is that it does so
through a  preheating mechanism \cite{steve2}. For the purpose of the 
present paper, the term `curvaton' covers all three cases.}
There is another aspect of the curvaton
hypothesis though, that has received hardly any attention so far.
This is the fact that the task of building a viable model of inflation
becomes much easier, if the model is liberated from the requirement 
that the inflaton be responsible for the curvatureu perturbation.

The layout of othe paper is as follows.
In Section \ref{sslorol}
 we recall the basics of slow-roll inflation. In Section \ref{sexpected} we
recall the expected form of the  potential in field theory. 
In Section \ref{slib} we
examine specific slow-roll models to see what is the effect of liberating
them. In Section \ref{sfasrol} we ask whether 
cosmological scales can instead leave the horizon during fast-roll inflation,
and in Section \ref{stherm} we ask the same question for thermal inflation.
We conclude in Section \ref{sconc}.
 Throughout the paper we use 
units such that \mbox{$\hbar=c=1$}, in which Newton's gravitational constant
is \mbox{$8\pi G=\mpl^{-2}$}, where \mbox{$\mpl=2.4\times 10^{18}\GeV$} is the
reduced Planck mass.

\section{Slow-roll inflation}

\label{sslorol}

\subsection{The basic equations}

Slow-roll inflation, with a single-component inflaton field, is described by 
the following basic equations   \cite{book,treview}. In these equations, 
 $\phi$ is the inflaton field  whose value determines the end of 
inflation,  and $V=V(\phi)$ is the  potential during 
inflation. The other quantities are  
 the scale factor of the Universe $a$, 
 the Hubble parameter  $H=\dot a/a$, and
  the wavenumber $k/a$ of the cosmological perturbations.

The potential satisfies  the  flatness  conditions
\bea
\epsilon&\ll &1 \label{flat1}\\
|\eta|&\ll& 1 \label{flat2} 
\eea
with $\epsilon\equiv  \frac12\mpl^2(V'/V)^2$ and 
$\eta\equiv  \mpl^2V''/V$, where the prime denotes derivative with respect to
 the inflaton field $\phi$. The inflaton's trajectory will then be
an attractor, satisfying the slow--roll approximation
\be
3H\dot\phi\simeq -V'
\,.
\label{sr}
\ee
The energy density is $\rho\simeq V$ leading to
\be
V\simeq 3\mpl^2 H^2
\label{vexp}
\,.
\ee
Given the potential, \eqs{sr}{vexp} can be integrated to give the
inflaton trajectory $\phi(t)$ up to the choice of $t=0$.

The inflaton generates a Gaussian curvature perturbation with spectrum
\be
\frac4{25}\calpr(k)
 = \frac1{75\pi^2\mpl^6}\frac{V^3}{V'^2}\left.{}\right|_*  \,.
\label{delh}
\ee
where the star denotes the epoch of horizon exit $k=aH$.
 \eq{sr}
determines the 
 number $N(k)$ of $e$-folds of slow-roll inflation occuring after 
horizon exit to be
\be
N(k) = \mpl^{-2} \int^{\phi_*}_{\phi\sub{end}} \( \frac{V}{V'} \) \diff \phi
\,.
\label{nofk}
\ee

On cosmological scales, $N(k)$ is given by
\bea
 N(k)
&=& 67 - \ln\frac{k}{H_0} -  \ln\(\frac{\mpl}{V_0^{1/4}}\) - \Delta \\
\Delta &\equiv & 
\frac13\ln\(\frac{V_0^{1/4}}{T\sub{reh}}\)  + N_0
\label{nefolds}
\,.
\eea
Here $H_0$ is the Hubble parameter at present, 
$T\sub{reh}$ is the reheating temperature after inflation, 
and $N_0=0$ if the Universe remains radiation-dominated after reheating,
until the onset of the present matter-dominated era.    The number
$N_0$ is positive if there is more inflation after the almost-exponential
era, or if radiation-domination is interrupted by  one or more
matter-dominated eras. There is no reasonable cosmology for which
$N_0$ is negative and {\em a fortiori} none for which $\Delta$ is negative.
Knowing $N(k)$ and $\phi\sub{end}$, \eq{nofk} determines $\phi_*$ and then
\eq{delh} determines  the curvature perturbation generated by the inflaton.

\subsection{The inflaton hypothesis}
According to the inflaton hypothesis,  the curvature perturbation 
on cosmological scales remains constant as long as these scales are
far outside the horizon.
Comparing \eqs{CMBnorm}{delh},
 this requires
at the epoch when the CMB scale leaves the horizon
the {\em CMB normalization},
\be
V^{1/4} = 0.027 \epsilon^{1/4} \mpl
\,.
\ee

Differentiating \eq{delh} and using the slow-roll expression
$3H\dot\phi=-V'$ gives the 
 spectral index 
\be
n= 1 + 2\eta - 6\epsilon
\,.
\label{ninf0}
\ee
(From now on, $\eta$ and $\epsilon$ will always be evaluated at horizon
exit.)
In nearly all 
inflation models, $\phi/\mpl$ is very  small while cosmological
scales leave the horizon, making $\epsilon$ completely negligible 
\cite{book,treview} so that
for practical purposes
\be
n= 1 + 2\eta 
\label{ninf}
\,.
\ee
In a large class of models, $\eta$ has the form
\be
\eta(\phi) = {\rm \, const \,} \phi^{p-2}
\ee
with $p<1$ or $p>2$. This leads  to
\be
n(k) - 1 = - \frac {p-1}{p-2} \frac 2{N(k)} < 0
\,,
\ee
making $n$ on on cosmological scales almost scale-independent 
and significantly  below 1. Within a few years $n$ will be determined
with an accuracy of order $\pm 0.01$, allowing this relation to be confronted
with observation  \cite{p00n1}. 

\subsection{The curvaton hypothesis}
In this paper we  adopt  the curvaton hypothesis,
that  the curvature perturbation comes primarily from the vacuum fluctuation
of some curvaton field different from the inflaton.
 Instead of the CMB normalization we therefore
have the {\em CMB bound},
\be
V^{1/4} \ll 0 .027 \epsilon^{1/4} \mpl
\,.
\label{CMBv}
\ee
Requiring that the curvature perturbation be, say, less than $1\%$ of the 
observed value, the left-hand-side of this expression must be less than
$10\%$ of the total giving
\be
V^{1/4} < 2\times 10^{15}\GeV
\label{v15}
\,,
\ee
which means that a gravitational wave signal will never be detected
in the CMB anisotropy \cite{gw02}.

In the curvaton model, the 
 condition  $|\eta_{\sigma\sigma}|\lsim  1$ is needed so that the
vacuum fluctuation of $\sigma$ is converted into a classical one, where
\be
\eta_{\sigma\sigma}\equiv \frac{\mpl^2}{V}
\frac{\partial^2 V}{\partial \sigma^2}
\ee
Taking $|\eta_{\sigma\sigma}|\ll 1$, the 
 spectral index in  the curvaton model is
 given by \cite{Lyth:2001nq}
\be
n = 1 + 2\eta_{\sigma\sigma} - 2\epsilon
\,,
\label{ncurv}
\ee
where the right hand side is evaluated at the epoch of horizon exit. 
Discounting an accidental cancellation, \eq{nbound} requires
\be
\epsilon \lsim 1/20 
\ee

In the large class of inflation models where $\epsilon$ is completely
negligible, $n-1$ is determined by $\eta_{\sigma\sigma}$ which specifies
the second derivative of the potential in the {\em curvaton} direction.
Since there is no reason why this quantity should increase during inflation,
it is reasonable to suppose that  it is negligible on cosmological scales,
leading to a spectral index indistinguishable from 1. 

\section{The expected form of the inflaton potential}

\label{sexpected}

According to the usual assumption (see Section \ref{smoving} for a possible
exception)  a single effective field theory holds
from the epoch of inflation until the present. To keep the Higgs stable
this field theory presumably respects supersymmetry (SUSY), which presumably
is  local  corresponding to supergravity 
(SUGRA). In the vacuum, SUSY is obviously broken, and it is also broken
in the early Universe because of the nonzero energy density.
The  breaking at the level of SUGRA must be  spontaneous,
which strongly suggests
$N=1$ SUGRA 
 since it seems very difficult to  spontaneously break
$N>1$ to the Standard Model. Although the detailed form of the SUGRA theory
is not known, certain features are expected \cite{wein,pol,book,treview}
on the basis of generic ideas about
field theory, and about the presumed underlying string theory involving extra
dimensions that have been integrated out. In this section we 
recall the main feature that are relevant for inflation model-building.

\subsection{The potential near the vacuum}

In the vacuum, phenomenology demands that
the sector of the theory
in which spontaneous breaking occurs  (the SB sector) be distinct from the
sector containing the 
Standard Model and its minimal extension (the MSSM). 
To a good approximation, the theory in the MSSM
sector should respect  global SUSY with explicit (soft) SUSY 
breaking terms. The SB sector may communicate with the
MSSM 
sector  by interactions which are  of gravitational strength
(gravity mediated SUSY breaking),  or stronger and typically involving
a gauge symmetry  (gauge, gaugino  mediated supersymmetry breaking). 
There is also the
case of anomaly-mediated SUSY breaking, where the effect of spontaneous SUSY
breaking is felt only through the gravitational anomaly. 

In the next section we are going to consider some supersymmetric models of 
inflation. In them, as in all known models,
 the inflaton 
 belongs  to a sector of the
theory which (at least during inflation)
is to some extent decoupled from the MSSM  sector.
Depending on the model, the SB sector
during inflation may or may not be the same as the SB sector
in the vacuum. Where it is different, the  possibilities for 
communication between the SB sector and the inflaton sector are the 
same as those already mentioned for the case of the vacuum (gravity-mediated
etc.). There is also the additional possibility that the SUGRA during inflation
is broken in the inflaton sector itself, with no separate SB breaking sector.
We shall encounter examples of all these cases.

The
 form of the scalar field  potential is
\be
V(\phi_i) = V_+(\phi_i) - 3\mpl^2 m_{3/2}^2(\phi_i)
\label{sugrapot}
\ee
where $\phi_i$ are the scalar fields.
Both terms are positive, and in the vacuum $m_\frac32$ is the gravitino mass. 
The first term $V_+$ is a measure of the
strength of spontaneous supersymmetry breaking generated by the potential.

Consider first the form of the potential near the vacuum, where $V$ 
(practically) vanishes.
The scale of supersymmetry breaking in the vacuum is denoted by $M\sub S$;
\be
\langle V_+ \rangle \equiv M\sub S^4
\ee
Since $V$ (practically) vanishes in the vacuum, the gravitino mass is given
by
\be
\sqrt 3 m_\frac32 = M\sub S^2/\mpl
\label{ginomass}
\,.
\ee
In the direction of a canonically-normalized real field $\phi$, let us for
the moment take $\phi=0$  as the vacuum value (VEV). Assuming for
simplicity a symmetry $\phi\to -\phi$, the  tree-level potential
will  have the form
\be
V(\phi) =  \frac12m^2\phi^2 + \frac14\lambda \phi^4 +
\sum_{{\rm even\,}i} \mpl^{4-i} 
\lambda_i \phi^i
\label{vvac}
\,.
\ee

For squark and slepton fields, the mass vanishes in 
the limit of unbroken SUSY,\footnote{The following is true for the Higgs fields
too, except that there is a small mass ($\mu$-term) in the limit of unbroken
SUSY.} and SUSY breaking gives a value
\bea
m &\sim&  \sqrt3 C M\sub S^2/\mpl \label{mvac} \\
& =& C m_\frac32 
\,,
\eea
where the number $C$ depends on the mediation 
from the hidden to the MSSM  sector;
\bea
C &\sim 1 & \ \ {\rm (gravity-mediated)} \nonumber \\
C &\gg 1 & \ \ {\rm (gauge/gaugino-mediated)} \nonumber \\
C &\sim 10^{-3} &  \ \ {\rm (anomaly-mediated)} 
\label{xval}
\,.
\eea
 To avoid detection and yet  keep the 
Higgs mass under control one needs $m\sim 100\GeV$ to $1\TeV$, giving 
\be
\sqrt C  M\sub S \sim 10^{10}\GeV
\,.
\ee
In a generic direction among the Higgs, squark and slepton fields,
$\lambda$ is of order 1 but there are `flat' directions where $\lambda$
is practically zero because it vanishes in the limit of unbroken SUSY.
It is generally assumed that $\lambda_i\sim 1$ even in the flat directions.

There may be other light fields, in particular moduli whose entire potential 
vanishes in the limit of unbroken SUSY. We shall make the usual assumption that
the potential of a modulus has the form
 $V=M\sub S^4 f(\phi/\mpl)$ with $f(x)$ and its 
derivatives of order 1 at a generic field value in the range $\phi\lsim \mpl$.
 This corresponds to $m\sim m_\frac32$ and
$|\lambda|\sim |\lambda_i|  \sim (m/\mpl)^2$.

\subsection{The potential during inflation}

During inflation, the potential $V$ is large so
at least one field must be far from the vacuum. If the only field with 
significant displacement is the inflaton we have non-hybrid inflation and
\eq{vvac} applies with $\phi$ the inflaton. Hybrid inflation is the case
where a second field is significantly displaced, being responsible for
$V_0$.
 We focus on slow-roll 
inflation, which requires
the flatness conditions \eqs{flat1}{flat2}.
One possibility (`chaotic inflation') is a quadratic or quartic
potential with $\phi\gg \mpl$, 
corresponding to extreme suppression of the relevant non-renormalizable
terms in \eq{vvac}. Barring this case, all proposed models of inflation 
suppose that the tree-level inflaton potential during inflation is
of the form 
\be
V(\phi) = V_0 \pm \frac12 m^2 \phi^2  + \sum_{i=3}^\infty
 \lambda_i  \mpl^{4-i} \phi^i
\label{vinf}
\,,
\ee
with $\phi\lsim \mpl$.
The origin is here taken to be a maximum or a minimum, the former  option
being mandatory for non-hybrid inflation. (Typically, the origin is
the fixed point of the relevant global and/or gauge symmetries.)
During slow-roll inflation, the 
 constant term $V_0$ dominates, 
and  only  one or two 
terms in the expansion are supposed to be
significant. Usually these are renormalizable terms
(quadratic, cubic or quartic). To achieve inflation over a range of $\phi$,
each term separately should satisfy \eq{flat2}, and then $\phi\ll\mpl$ ensures
that \eq{flat1} is satisfied. 

During inflation, SUSY is broken by the energy density $\simeq V_0$, and
to keep the potential as flat as possible one assumes that the relevant
$\phi$-dependent terms in \eq{vinf} come entirely
 from the SUSY breaking.  
(If global SUSY is a 
good approximation this can be achieved by choosing $\phi$ to be  a flat 
direction. Alternatively it can be achieved by choosing $\phi$ to be a 
modulus.)
This, however, is not enough to guarantee \eq{flat2}.
On the contrary, for
a {\em generic} field  during inflation, the mass generated by SUSY breaking
is at least\footnote
{The mass is  bigger for gauge/gaugino-mediated case, making that case 
definitely unsuitable for the inflaton. In contrast with the
vacuum case, the mass during inflation
is not generically smaller in the anomaly-mediated case
(corresponding to no-scale SUGRA) \cite{cllsw,glm}. 
We are assuming that SUSY is broken
by $F$ terms, but breaking it by $D$ terms may  not  help. This is 
because \cite{km} the loop correction then requires that
inflation  takes place at $\phi\sim\mpl$,  making it difficult to control
 the non-renormalizable terms.}
\be
|m^2| \sim V_+/\mpl^2
\label{mest}
\,.
\ee
If there were a significant degree of cancellation between the
 two terms of \eq{sugrapot}, this estimate  would  strongly violate \eq{flat2}.
Such a cancellation would be expected if $V_0\ll M\sub S^4$, and accordingly
such low values of $V_0$ are disfavoured (see Sections \ref{sdineriotto}
and \ref{sking}). For bigger 
values of $V_0$, the estimate \eq{mest} violates \eq{flat2}
 only marginally. 
A sufficiently small mass can in this case occur through
and accident, through
 a running mass, through a special form for the potential during inflation,
or because of 
a global symmetry \cite{treview}.

\section{Liberated models of slow-roll inflation}

\label{slib}

In this section we examine some models of
slow-roll inflation. Under the inflaton hypothesis,
reviews of most of them  have been 
given already \cite{book,treview},
 to which we refer for further detail
and references. We omit some models which have too many parameters for a 
definite statement to be made about the effect of liberating them, 
and some proposals which were not carried
through to the point of producing a specific model, but otherwise our coverage
of extant models is fairly comprehensive.

\subsection{Slow-roll modular inflation}
\label{modular}

Starting with \cite{bg}, many authors 
(eg.\ \cite{modular1,modular2,natural2})
have considered the possibility of
slow-roll 
non-hybrid inflation where the inflaton is a modulus with the kind of potential
we described earlier. Inflation would take place around a maximum, where
\be
V=V_0-\frac12m^2\phi^2 +\cdots
\,,
\label{invquad}
\ee
with the remaining terms negligible in the regime $V\sim V_0$
and
\be
m^2\sim V_0/\mpl^2
\,,
\label{modmass}
\ee 
which will also be
roughly the mass-squared of $\phi$ in the vacuum at $\langle \phi \rangle
\sim \mpl$.

In the regime $\phi\ll \mpl$ where \eq{invquad} is valid, the
flatness parameter $\epsilon$ is practically zero but the other
flatness parameter $\eta$ is given by
\be
\eta\simeq - \frac{m^2\mpl^2}{V_0} \simeq -\frac{m^2}{3H^2}
\label{etamodular}
\,.
\ee
The generic estimate \eq{modmass} would make $|\eta|\sim 1$,
but one can imagine that the actual value of $\eta$
calculated for some
modulus in a specific string theory is significantly smaller.
This is the hypothesis of slow-roll modular inflation.

Inflation with the  potential  \eq{invquad} (with the remaining terms
negligible in the regime $V\simeq V_0$)
 has also been considered without any specific hypothesis about the
identity of the inflaton (`natural' \cite{natural2,natural1} and`topological' 
\cite{topological} inflation). The 
 CMB bound   for such a  potential is \cite{treview}
$V_0^{1/4}\ll  10^{16}\GeV$, corresponding to mass 
\be
m\ll  10^{14}\GeV
\label{modbound}
\,.
\ee

With $\phi$ identified as a modulus, it is usually supposed that $m$
is of order the gravitino mass which is expected to be at most of order
$100\TeV$ (corresponding to anomaly-mediated \cite{anmed}
supersymmetry breaking). This is in strong conflict with the
inflaton hypothesis, according to which the CMB bound is
saturated. 
In contrast, if we adopt the curvaton hypothesis there is no problem.

It has been proposed  instead 
\cite{modular2} that the modulus mass during inflation
is of order $10^{14}\GeV$ so that 
 the inflaton
hypothesis is satisfied.
However,  in all models  invoking the potential \eqs{invquad}{modmass}
the inflaton  hypothesis  presents another  possible 
problem.
This is the fact that even if $|\eta|$ is suppressed sufficiently to allow
slow-roll inflation, the  spectral index
$n\simeq 1+2\eta$  may 
turn out to be too far below 1 to be compatible with observation.
This problem cannot occur under
 the curvaton hypothesis,  since the 
spectral index \eq{ncurv}
 is then
independent of $\eta$.

We mention in passing the existence of a quite different model,
which  also gives $n$ too far below 1 on the inflaton hypothesis
but is liberated by the curvaton hypothesis.
It comes up  as an example of a model  \cite{Cohn:2000hc} in which 
the flatness of the inflaton potential in hybrid inflation is 
protected by a non-Abelian global symmetry, broken only by non-renormalizable
terms. The example, the only one worked 
out in detail so far \cite{Stewart:2000pa}, gives $\epsilon$ very small but
 a spectral index too far below 1 on the inflaton hypothesis.
(The spectral index on the inflaton hypothesis 
is actually given by a more complicated  formula than 
\eq{ninf} in this model, 
because  the inflaton in this model has two components.)

\subsection{The gauge--mediated model of Dine and Riotto}

\label{sdineriotto}
In the previous model the mass-squared term is supposed to dominate
until almost the end of inflation. An alternative is to have
the mass-squared
term dominating while  cosmological scales leave the horizon, but to have
a $-\lambda\phi^4$  term coming in well before the end of inflation.
If $m$ and $\lambda$ are regarded as free parameters this represent 
fine-tuning, but we shall now consider a model in which the parameters
are related so as to produce the inflaton evolution that we just described.

The  model \cite{Dine:1997kf} works in the approximation of global
supersymmetry, and it 
assumes that SUSY breaking in the MSSM sector is gauge-mediated
according to the following standard scheme.
In a secluded sector, a field $X$ and its auxiliary field $F_X\equiv 
\partial W/\partial X$ (with $W$ the superpotential)
are both supposed to have nonzero VEVs, the latter determining the
scale of supersymmetry breaking,
\be
\langle F_X \rangle \equiv M\sub S^2
\ee
One requires
 $\Lambda\equiv F_X/\langle X \rangle$ of order $10^5\GeV$,
 so that the radiatively-generated soft 
masses in the MSSM  sector have the desired magnitude $\sim 10^2\GeV$.

The inflaton field $\phi$  in this model is the real part of a 
gauge singlet field $S$, and its VEV generates
the $\mu$ term of the MSSM \cite{Dine:1995ag}.
During inflation, $X$ and $F_X$ as well as the gravitino mass 
are supposed to be close to their vacuum values, 
implying some degree of cancellation between the
two terms  of \eq{sugrapot}. 

The superpotential is 
\be
W= - \frac{\beta XS^4}{\mpl^2} + \frac {S^5}{\mpl^2}
+ \mpl^{-n} S^{n+1} H_U H_D + \cdots \,.
\label{dinew}
\ee
This  structure can be enforced
by discrete symmetries.
We have exhibited 
 a coefficient $\beta\lsim 1$ 
though eventually we shall favour a value of order 1.
All other coefficients are assumed to be of order 1 from the outset,
and throughout the calculation we shall ignore numerical factors of order 1.
The third term of the superpotential generates the $\mu$
term of the MSSM, but plays no role during inflation. 

The dots represent the contributions to $W$ that do not involve
$S$. They generate, among other things, a contribution to $F_X$,
 which is assumed to be  the dominant contribution.
 Since we are assuming that $F_X$ is close to its VEV during inflation,
this requires  for consistency 
\be
\beta \langle \phi^4 \rangle \ll \mpl^2 M\sub S^2
\label{consistency}
\ee

The  corresponding potential is
\be
V \simeq V_0 -m^2\phi^2 - \frac14 \lambda \phi^4 + \( \mpl^{-4}{\phi^8}
+ \beta^2 \mpl^{-4} X^2 \phi^6 \) +\cdots
\label{veq}
\ee
with
\be
\lambda = \beta \mpl^{-2} M\sub S^2 
\ee
and
\be
m^2 \equiv \alpha V_0/\mpl^2 \,.
\ee
(Note that the slow--roll condition \eq{flat2} requires $\alpha\ll 1$.)
We  have dropped a term which is significant only if the two terms in 
the bracket have a similar magnitude.

The VEV $\langle \phi\rangle$  is  determined by minimizing this
 potential.\footnote
{The following results are taken from \cite{treview}, correcting the
erroneous treatment of \cite{Dine:1997kf} which misses the second of the two
 cases below.}
In the case $\beta M\sub S^2\lsim (\Lambda^2 \mpl)^{2/3} \sim ( 10^9\GeV)^2$,
one finds 
$\langle \phi^4 \rangle \sim \beta M\sub S^2 \mpl^2 $ 
(marginally consistent with
\eq{consistency}~) and 
\be
V_0 \sim \beta^2 M\sub S^4
\label{pot1}
\ee
In the opposite case $M\sub S^2\gsim (10^9\GeV)^2 $ one finds instead
$\langle \phi \rangle ^2 \sim \mpl^2\Lambda^2 /\beta M\sub S^2$ (comfortably
 consistent with
\eq{consistency}~) and
\be
V_0 \sim \(\frac{\Lambda^4\mpl^2}{\beta^3  M\sub S^6}\)  \beta ^2 M\sub S^4
\label{pot2}
\ee

During inflation the bracketed term in \eq{veq} is negligible,
while  both of the first two terms are in general significant. A
careful calculation reveals   \cite{treview} that the CMB bound is
in all cases $\lambda\ll 10^{-15}$, corresponding to
\be
\sqrt{\beta M\sub S^2} \ll  10^{10.5} \GeV
\ee
On the inflaton hypothesis  the bound is saturated, which 
practically kills the model  for two  reasons.
First, gauge--mediated supersymmetry breaking  requires
$M\sub S$ to be significantly  less than $10^{10}\GeV$, to validate the
assumption that gravity--mediated breaking is negligible. This is in mild
conflict with the expectation $\beta\lsim 1$.
 Second, even with $\beta\sim 1$,
\eq{pot2} requires $V_0\sim 10^{-7} M\sub S^4$ which means that the
 two terms of \eq{sugrapot} must cancel with an accuracy $10^{-7}$,
 and means also that
 $\alpha$ has to be suppressed by a more than a 
factor $10^{-7}$ below its natural value.

Both  of these problems disappear if we adopt the curvaton hypothesis
while retaining $\beta\sim 1$. In particular,  the only requirement on
$\alpha$ is that it be small enough for slow-roll, implying 
only mild tuning of $\alpha$ below the natural value of order 1.

\subsection{Constraints on hybrid inflation}

\label{sconstraints}
Now we turn to hybrid inflation models. Before considering a couple
of specific models, we consider the constraints on the parameter space
\cite{Lyth:1999ty}
arising from the fact that hybrid inflation necessarily involves an
interaction between the inflaton field $\phi$ and some other field $\chi$, 
which
in turn generates a loop correction that might 
violate the flatness conditions.

The potential for hybrid inflation is basically of the form
\be
V(\phi,\chi)= V_0 + \Delta V(\phi) 
-\frac12\mpsis\chi^2 +\frac12{\lambda'}\chi^2\phi^2
+ \frac14\lambda\chi^4 \,.
\label{fullpot}
\ee
(Modifications of the last two terms are sometimes considered, which 
typically do not affect the following considerations.)
Inflation takes place in the regime $\phi^2>\phi\sub c^2$, where
\be
\phi\sub c \equiv |m_\chi|/\sqrt{\lambda'} \,.
\label{phic}
\ee
In this regime, $\chi$ vanishes 
and the inflaton potential is
\be
V=V_0+ \Delta V(\phi) \,. \label{vofphi2}
\ee
The constant term $V_0$ is assumed to dominate during inflation.

The last term of \eq{fullpot} serves only to determine the 
VEV of $\chi$, achieved when $\phi$ falls below
$\phi\sub c$. Using
that fact that $V_0$ vanishes in the vacuum, one learns that
the VEV is
\be
\langle\chi\rangle=2 V_0^{1/2}/|m_\chi| \,, 
\label{mofv} \ee
and that
\be
\lambda=\frac{4V_0}{\langle\chi\rangle^4} = \frac{m_\chi^4}{4V_0} \,.
\label{lamofv}
\ee

The main difference between hybrid inflation models is in the form of
$\Delta V$.
In the original model  \cite{Linde:1991km},
\be
\Delta V(\phi) = \frac12m^2\phi^2 \,.
\label{mterm}
\ee
Later, models were proposed where $\Delta V$ is instead dominated by
the loop correction coming from the interaction of the inflaton with
$\chi$ and its superpartner.
These models were formulated in the approximation of 
spontaneously broken global SUSY (involving either an $F$ term
\cite{cllsw,dss} or a $D$ term \cite{ewan,bd,halyo}) giving a loop
correction of the form
 \be
 \Delta V =V_0 \frac{g^2}{8\pi^2}\ln \frac\phi Q 
\label{sponloop}
\,,
\ee
 with the renormalization scale $Q$ chosen to make the second term small.
The coupling $g$ is of order 1 for $D$-term inflation (gauge coupling) but
may be smaller for $F$-term inflation (Yukawa coupling).

Alternatively, it may be that {\em softly} broken global SUSY is a good
approximation during inflation. In that case the loop correction, coming
from $\chi$ and its superpartner or from any other supermultiplet, giving
 \be
\Delta  V =  f^2  \phi^2 \ln \frac\phi Q 
\label{softloop}
\,,
\ee
where $f$ is a measure of the coupling strength.
This is the starting point for the running-mass
model of Section \ref{srun}.

Now we come to the source of the constraint on the parameter space of
hybrid models. 
The  loop correction coming from 
$\chi$ and its superpartner will presumably not be accurately canceled
by other terms over the relevant range of $\phi$.
The contribution of this loop 
 provides an approximate
 lower bound
for $\Delta V$ and $\Delta V'$,
and the constraint on the parameter space comes from the requirement that this
bound should respect the flatness conditions \eqs{flat1}{flat2}.
This constraint has been evaluated in \cite{Lyth:1999ty}. We consider only
the weakest constraint,  corresponding to \eq{sponloop}.
Under the inflaton hypothesis,  
 it  may be written in the form
\be
\langle\chi\rangle^4 \phi\sub{\sc CMB} \gsim \( 10^9\GeV \)^5\,, 
\label{advert}
\,,
\ee
where the subscript CMB refers to the epoch at which the scales explored
by the CMB anisotropy leave the horizon.
This  becomes an explicit constraint on the parameter space,
once a form for $\Delta V$ is specified which allows $\phi\sub{\sc CMB}$ to
be calculated. For example, with the original
form \eq{mterm},
\bea
\langle \chi \rangle^3 &\lsim & 5\times 10^{-5} \sqrt {\lambda'} \mpl^3\\
\lambda' &\lsim &(\eta/22)^{3/2} \\
\eta &\lsim & (90 \langle\chi\rangle\mpl )^4 
\label{mbound1}
\,,
\eea
where $\eta=m^2\mpl^2/V_0$.

These constraints 
 are very powerful. In particular, \eq{advert} means
that  un--liberated
hybrid inflation cannot be expected to work if the ultra--violet cutoff
is below $10^9\GeV$ (since both $\langle \chi \rangle$ and $\phi\sub{\sc CMB}$
are expected to be below this value).
In particular, one cannot expect un--liberated hybrid inflation to work if
there is an extra dimension with size $\gsim (10^9\GeV)^{-1}$.

In contrast, for liberated hybrid inflation
 \eq{advert} becomes
\be
\langle \chi\rangle^2 \phi\sub{\sc CMB} \gsim \( 10^4\GeV \) ^3 
\sqrt{N\sub{\sc CMB}}
\label{advert2}
\ee
which in the case of \eq{mterm} becomes
\be
\lambda' V_0^{1/2}  \lsim 16\pi^2 m \langle \chi \rangle
\ee
These constraints are far weaker, and in particular
\eq{advert2} makes liberated hybrid inflation feasible for
an ultra--violet cutoff as low as $10^4\GeV$.

\subsection{Running-mass hybrid inflation}

\label{srun}
The running-mass model \cite{running,Covi:1998mb} assumes that the inflaton 
sector can be described by global SUSY with explicit (soft) breaking,
as opposed to the spontaneous breaking that gives \eq{sponloop}.
The inflaton mass is supposed to run significantly, corresponding to a
 gauge or Yukawa interaction between the 
inflaton and some supermultiplet (in the case of a Yukawa, the interaction
might be the one involving $\chi$). At some high scale
 the inflaton mass-squared is supposed to 
 have the generic magnitude of order $V/\mpl^2$. At some lower scale
$Q$
the running mass is supposed to pass through zero, generating a
maximum or minimum of the potential at some field value $\phi_*
\sim Q$.  Near this value the potential
is well-approximated  \cite{p00n1} by the 1-loop correction \eq{softloop}.
The resulting potential may be written
\be
V = V_0 \left\{1 - {c\over 2} \frac{\phi^2}{\mpl^2} \[ \ln(\phi/\phi_*)
-\frac12 \] \right\}
\,.
\ee
The CMB bound for this model is
\be
\frac{V_0^{1/2}}{\mpl} \ll 2\times 10^{-5} 
c\,\phi_{\sc CMB}
\left|\ln\left(\frac{\phi_*}{\phi\sub{\sc CMB}}\right)\right|
\ee

Under the inflaton hypothesis, the 
spectral index predicted by the running mass model is
\bea
n(k)-1 &=& 
2c\left\{\ln\!\left(\frac{\phi_*}{\phi\sub{\sc CMB}}\right)
\exp [c \Delta N(k)]
 - 1\right\} 
\label{nofsc}
\\
\Delta N (k) &\equiv& N(k\sub{\sc CMB})- N(k)
\,.
\eea
The requirement that $n(k)$ be within observational bounds
even now constrains the parameter space \cite{p00n1,p00n2,p02laura,clmprep}, 
and in  the  future it might rule out the model altogether.
Saturating the CMB bound may also
 make the value of $\phi_*$ rather  low compared with 
theoretical expectations \cite{p02laura,clprep}. Both of these possible 
problems disappear
if the model is liberated.

\subsection{The hybrid inflation model of Bastero-Gil and King}

\label{sking}
Like the model of Section \ref{sdineriotto}, 
this one assumes that supersymmetry is broken during
inflation by the same mechanism as in the vacuum, at the same scale $M\sub S$
 \cite{Bastero-Gil:1997vn}.
In contrast with that model though, supersymmetry breaking is supposed to be
transmitted with only gravitational strength, both to the inflaton and to
the MSSM sector; in other
words, we are dealing exclusively 
with gravity--mediated supersymmetry breaking, and $M\sub S\sim
10^{10}\GeV$.

In this model, the inflaton generates the $\mu$ term
only indirectly. 
The  superpotential is
\be
W = \lambda {\cal N} H\sub U H\sub D - \kappa S {\cal N}^2 +\cdots
\ee
where  $\cal N$ and $S$ are gauge singlets. They respect the
Peccei-Quinn global symmetry, whose pseudo-Goldstone boson is the axion
which ensures  the CP invariance of the strong interaction. 

The axion is practically massless, and can be chosen such that $S$ is real.
 The inflaton is the canonically--normalized quantity
$\phi=\sqrt{2}\;{\rm Re}\,S$. During 
inflation $H\sub U H\sub D$ is negligible. Writing $\sqrt 2\,{\cal N}=
{\cal N}_1 + i {\cal N}_2$,
and including a soft supersymmetry breaking trilinear
term $2A \kappa\phi {\cal N}^2 + {\,\rm c.c}$
(with $A $ taken to be real) as well as soft supersymmetry breaking 
mass terms, the potential is
\be
V=V_0+ \kappa^2 |{\cal N}|^4+ \frac12\sum_i m_i^2(\phi){\cal N}_i^2+
\frac12 m_\phi^2 \phi^2,
\ee
where 
\bea
m_1^2(\phi) &=& m_1^2 -2 \kappa A \phi + 4 \kappa^2 \phi^2 \,, \label{n1m}\\
m_2^2(\phi) &=& m_2^2 + 2 \kappa A \phi + 4 \kappa^2 \phi^2 \,.
\label{n2m}\\
\eea

The soft supersymmetry breaking parameters $m_i$  and 
$A$ are supposed to have the typical values for gravity--mediated
supersymmetry breaking, 
\be
m_i \sim A \sim M\sub S^2/\mpl (\sim 100\GeV) 
\ee
In contrast, in order to achieve slow--roll inflation, the mass $m_\phi$
is supposed to be
\be
m_\phi^2 \sim \alpha V_0/\mpl^2
\ee
with $\alpha\ll 1$.

The VEVs are given by
\bea
\langle \phi\rangle &=& \frac{A}{4\kappa}, \\
\langle {\cal N}_1\rangle &=& \frac{A}{2\sqrt 2 \kappa}\sqrt
{1-4\frac{m_1^2}{A^2}} \\
\langle {\cal N}_2\rangle &=& 0 \,.
\eea
where we ignored the tiny effect of $m_\phi$.
It is assumed that $4m_1^2$ is somewhat below
$A^2$, so that 
\be
A \sim \kappa\langle {\cal N}_1\rangle \sim \kappa
\langle \phi\rangle \sim 1\TeV \,.
\ee
To have the VEVs at the axion scale, say $10^{13}\GeV$,
we require $\kappa\sim 10^{-10}$. Also, $\lambda$ should have a 
similar value, since $\lambda \langle {\cal N}_1\rangle$ will be the 
$\mu$ parameter of the MSSM. 
The tiny couplings $\kappa$ and $\lambda$ are supposed to be 
products of several
terms like $(\psi/\mpl)$ where $\psi$ is the VEV of a field that is 
integrated out.

During inflation, the fields ${\cal N}_i$ are trapped at the origin,
and 
\be
V=V_0 +\frac12m_\phi^2\phi^2 \,.
\ee
The field ${\cal N}_1$ is destabilized if $\phi$ lies between the values
\be
\phi\sub c^{\pm} = \frac{A}{4\kappa}\( 1\pm \sqrt{1-4\frac{m_1^2}{A^2} }
\)
\sim A/\kappa
\,.
\ee

If $m_\phi^2$ is positive the model gives ordinary hybrid inflation
ending at $\phi\sub c^+$, but if it is negative it gives inverted 
hybrid inflation ending at $\phi\sub c^-$. 
The height of the potential is 
\be
V_0^{1/4}\sim A/\sqrt \kappa \sim 10^8\GeV
\label{vheight}
\ee
As $V_0$ is a factor $10^{-8}$ below $M\sub S^4$, the flatness condition
$\alpha\lsim 1$ requires that the $|m_\phi|^2$ is more than a factor
$10^{-8}$ below the generic value
given by \eq{mest}. 

The CMB bound for this model is
\be
A\ll  5\times 10^{-4} \alpha \, e^{\pm \alpha N_{\sc CMB}} \mpl 
\ee
On the inflaton hypothesis,  the bound is saturated, corresponding to
$\alpha \sim 10^{-12}$ which requires that $|m_\phi|^2$ is a factor
$10^{-12-8}=10^{-20}$ below its generic value. Adopting instead the
curvaton hypothesis, we require only the milder suppression by a factor
$10^{-8}$ required by the flatness condition. This suppression comes from
the mismatch between the height of the potential $V_0^{1/4}\sim 10^8\GeV$
 and the supersymmetry breaking scale $M\sub S\sim 10^{10}\GeV$. It would be
interesting to see if the model could be modified to reduce this 
mismatch, for instance by
lowering  $M\sub S$ and generating $m_i$ and $A$ 
 through interactions
with the supersymmetry breaking sector which are of more than gravitational
strength.\footnote{After the above words were written,  a 
modification of the model has been formulated \cite{steve2}
doing just that. In this model the Higgs field is responsible for the
curvature perturbation, through a modified version of the curvaton mechanism.}

\subsection{Inflation from a moving brane}

\label{smoving}
In all of the models considered so far, the Universe is supposed to be 
described by a single effective field theory
from inflation until the present day. In particular, this field theory
is supposed to describe the reheating process that replaces the inflaton
field with thermalized radiation.

As we have seen, the assumption of a single field theory makes it
quite difficult to keep the inflaton potential flat enough.
Now we come to a proposal which at least in spirit is different.
This is the proposal \cite{Dvali:1998pa,Alexander:2001ks,quevedo,Dvali:2001fw,Shiu:2001sy}
that  the inflaton corresponds to the distance
between two branes moving in $d$  extra dimensions, reheating occuring when
the branes collide. At least in the current
implementations of this proposal, the $N=1$ supergravity that is presumed to
hold for the field theory containing the Standard Model is not presumed to
hold during inflation, and as a result it is 
 relatively easy to keep the potential sufficiently flat.
One interpretation of this state of affairs might be
that the field theory during inflation is different from the one after
reheating, both theories breaking down during reheating.

The form of the potential depends on the setup in the context of string 
theory.
In the original proposal \cite{Dvali:1998pa},
\be
V\simeq V_0(1- e^{-q\phi/\mpl})
\label{another}
\ee
with  $q$ of order 1. The scale $V_0^{1/4}$ of the potential is of order
the higher--dimensional Planck scale $M$, related to the size  $R$
of the extra dimensions by $R^d\sim \mpl^2/M^{2+d}$, and the minimal value
$M\sim \TeV$ was taken to be the favoured one.
The CMB bound for the  potential \eq{another} is \cite{treview}
\be
V_0^{1/4} \lsim 7\times 10^{15}\GeV
\ee
As the authors noted, this places $M$  far above the $\TeV$.
Liberating the model allows instead $M\sim \TeV$.

Later authors \cite{quevedo,Dvali:2001fw,Shiu:2001sy} 
have considered in more detail the
form of the potential to be expected on the basis of string theory,
finding in different regimes potentials such as \cite{Dvali:2001fw}
\be
V=V_0\(1 - \frac{V_0}{32\pi^2 \phi^4} \)
\ee
and \cite{quevedo}
\be
V=V_0 -\lambda \phi^4
\ee
Both of these forms have been proposed earlier on the basis of 
purely four--dimensional field theory, and their predictions
for the  CMB bound are  well known
\cite{treview}. 
By relating the parameters to the higher--dimensional quantum
gravity scale $M$, it is again found that the latter needs to be far 
above the $\TeV$ scale. Again, liberating the 
 models allows instead $M\sim \TeV$.

\begin{table}
\centering
\begin{tabular}{lll}
\hline
Type  & $|$ & The effect of liberation \\ \hline
Modular  & : & Allows the usual  modulus mass $\sim \TeV$ \\
Dine/Riotto & : & Removes extreme fine--tuning \\
Hybrid (generic) & : & Allows extra dimension $\sim (10^4\GeV)^{-1}$\\
Running mass & : & Removes danger from future CMB measurements\\
Bastero-Gil/King & : & Removes of alleviates extreme fine--tuning \\
Moving brane & : & Allows  extra dimension $\sim (10^3\GeV)^{-1}$ \\
\hline
\end{tabular}
\caption{Models of slow--roll inflation  which benefit from liberation}
\end{table}

\subsection{Models which do not benefit from liberation}

\label{snot}

\paragraph{\boldmath $D$-term inflation}

We end with four examples where liberation does nothing to improve the model.
 The first case is that of `$D$--term inflation' \cite{ewan,bd,halyo}  and 
a related $F$--term model \cite{cllsw,dss} leading to
 hybrid inflation dominated by the loop correction with
 spontaneously broken supersymmetry. The potential is given by \eq{sponloop},
and the  CMB bound is
 \be
 V^{1/4} \ll 6.0 \(\frac {50}{N} \)^{1/4} g \times 10^{15}\GeV
 \ee
 In the case of $D$--term inflation, even the upper bound is difficult to
 reconcile with the expectation from weakly coupled heterotic string theory.
 Liberating the model obviously does nothing to improve that situation. Nor
 does it remove another problem of $D$-term inflation,  
that the value  $g\sim 1$ corresponding to a gauge coupling leads to a
value of
 $\phi$ during inflation is of order $\mpl$ making the needed
 suppression of non--renormalizable terms difficult to understand. 
In the case of $F$-term inflation, $g$ is a Yukawa coupling which can be
small, possibly removing the second problem, and also the 
inflation scale is arbitrary as opposed to being tied to string theory.
Liberating the $F$-term model has therefore a neutral effect on both counts.

 \paragraph{Monomial potential (`chaotic inflation')}

 For the  $V=\frac12m^2\phi^2$
 the CMB bound is
 \be
 m\leq 2\times 10^{13}\GeV
 \,.
 \ee
 During inflation $\phi$ is much bigger than $\mpl$, making it again difficult
 to understand the absence of non--renormalizable terms. Liberating the model
 does not help with that problem, and has the unfortunate effect of
 removing its prediction that the primordial gravitational waves will  be 
 observable in the foreseeable future through the CMB anisotropy.

\paragraph{Extended inflation}

Extended inflation \cite{book,La:za}
gives in its simplest form a potential $V=V_0\exp(-\sqrt{2/p}\phi/\mpl)$,
leading to $\epsilon=1/p$ and $\eta=2/p$. In this model the end of inflation
occurs through bubble formation, and to keep the bubbles invisible on the
microwave sky requires $p< 10$ or $\epsilon > 0.1$. Under the inflaton 
hypothesis this gives spectral index $n<0.8$ which is too low compared with
the observational bound \eq{nbound}. However, liberation  probably
does not help
(in contrast with the situation for modular inflation) because the 
model has $\epsilon\gsim 1/10$. Unless there is cancellation between the
terms of \eq{ncurv}, this again makes $n\lsim 0.8$.

 \paragraph{Inflation from the trace anomaly}
 
Before the term `inflation' was coined, Starobinsky \cite{Starobinsky:te} 
proposed that effective slow--roll inflation could be caused 
  by higher derivative curvature terms, without an explicit scalar field.
 The proposal has recently been reexamined by Hawking et al. 
\cite{Hawking:2000bb},
 who estimate that in their version of the model
 the CMB bound is 
 \be
{N}_{\rm S}^2 \( 250 + 240\beta - 40\alpha \) \gg 10^{13}
 \ee
 where $\alpha$ and $\beta$ are the coefficients of higher--order curvature
 terms and ${N}_{\rm S}$ is the number of scalar fields.
 This has the unpleasant feature that at least one of the three quantities
 must be exponentially large even in the un-liberated cases, and liberating
 the model clearly does not help.




\section{Fast-roll inflation}\label{sfasrol}

In this section we return to inflation with the potential given by
\eqs{invquad}{modmass}, but now suppose the flatness parameter 
$\eta$ has magnitude of order 1 in accordance with the generic expectation
of  \eq{modmass}. Since the flatness condition $|\eta|\ll 1$ is now violated,
the slow-roll condition \eq{sr} is also violated. However, there can still
be inflation, which has been called fast-roll inflation \cite{Linde:2001ae}.
The exact classical equation for the inflaton field is
\be
\ddot \phi + 3H\dot\phi + V' = 0
\label{phieq}
\ee
Taking $V'=-m^2\phi$ from \eq{invquad} and making the approximation
$H=$\,constant, the solution of this equation is 
\be
\phi\propto e^{FHt}
\ee
where
\be
F\equiv \frac32 \( \sqrt{1+\frac43 |\eta|} - 1 \)
\label{F}
\ee
The approximation $H=$constant is justified because this solution 
gives 
\be
\epsilon \simeq -\frac{\dot H}{H^2}
=\frac12 \frac{\dot\phi^2}{\mpl^2H^2} = \frac12 F^2
\frac{\phi^2}{\mpl^2} \ll 1
\,.
\ee

Under the inflaton hypothesis, fast-roll inflation is not viable
because the predicted spectral index  \cite{Stewart:1993bc}
$n-1=2\eta + \frac23 \eta^2$ will almost certainly 
be too far from unity.\footnote
{For $\eta=-3$ the two terms cancel, but
 this  represents fine-tuning and in any case 
the neglected terms in \eq{invquad} will almost certainly change the
result and prevent the cancellation from occuring over the whole 
cosmological range of scales.}
Under the curvaton hypothesis this
 problem with the spectral index does not arise. We must ask, though,
whether enough $e$-folds of inflaton can be generated by fast-roll inflation.

The number of $e$-folds of inflation is $N=F^{-1}\ln(M_0/\phi_0)$
where $M_0\sim\mpl$ is the VEV of $\phi$ and $\phi_0$ is its initial
value. There is a lower bound on the  latter, coming from the requirement
that the classical motion of the field in a Hubble time, $H^{-1}\dot\phi$,
be bigger than the quantum fluctuation  $H/2\pi$. Since we are dealing with
$\eta\sim 1$, this requirement amounts to $\phi\gsim H$, giving
 \cite{Randall:1995dj}
\be
N\sub{roll} \simeq \frac1 F \ln\( \frac{M_P}{m} \)
\simeq \frac 2 F \ln \( \frac{\mpl}{V_0^{1/4}} \)
\label{nroll}
\,.
\ee

The minimum number of $e$-folds that are needed is given by \eq{nefolds}.
Setting $N_0=0$, 
we can use the inequalities $T\sub{reh} >10\MeV$
(from nucleosynthesis) and \mbox{$V_0^{1/4}<10^{15}\GeV$} 
(from \eq{v15}) to find
\be
\Delta < 15
\label{Dbound}
\ee
To ensure that there is no 
excessive quadrupole contribution to the CMB anisotropy 
(Grishchuk-Zeldovich effect) the 
 spectrum of the curvature perturbation
should extend down to comoving wavenumber \cite{book}
  $k\sim 10^{-2}H_0$.
 This is the biggest  cosmological scale,
which leaves the horizon at 
\be
 N\sub{big}
\simeq 72 -
\ln\(\frac{\mpl}{V_0^{1/4}}\)- \Delta
\,.
\label{nbig}
\ee
The smallest cosmological scale presumably is the one
 enclosing  a mass
of order  $10^6\msun$, corresponding  to $k\sim 10^{-5.5}H_0$.
It  leaves the horizon about 17 $e$-folds after the biggest
cosmological scale.

Requiring $N\sub{big}< N\sub{roll}$ gives
\be
|\eta|\lsim 2 \( \frac{X}{72-\Delta-X}\right) +
 \frac{4}{3} \( \frac{X}{72-\Delta-X}\right)^2
\,,
\ee
where $X\equiv \ln(\mpl/V_0^{1/4})$. 
For \mbox{$V_0^{1/4}\sim 10^{11}$GeV} 
and $\Delta=0$ the above gives \mbox{$|\eta|\sub{max}\approx 0.75$}. This 
can be somewhat relaxed if \mbox{$\Delta\neq 0$}, with
\mbox{$|\eta|\sub{max}(\Delta=15)\approx 1.09$}.


We conclude that fast-roll inflation {\em can} provide enough $e$-folds,
provided that the inflaton starts out very close to the origin.
In the next section, we will see how this condition may be achieved
by coupling the inflaton to a 
thermal bath existing before
inflation, which leads us to investigate thermal modular inflation below.
As we will show, in the case of thermal modular inflation one also has the 
extra bonus of somewhat relaxing the bound on $|\eta|$.

\section{Thermal inflation}

\label{stherm}

The inflation models that we looked at  in the previous section
 all involve an inflaton
field. We end by looking at thermal inflation 
\cite{bg,thermal1,thermal2,thermalrest,hs}.
Thermal  inflation is maintained by a finite-temperature correction
to the potential, and it ends when the temperature $T$
 falls below some critical
value. There is no inflaton field during thermal inflation, and 
all previous authors have therefore assumed
that the curvature perturbation originates during an earlier era of slow-roll
inflation, with perhaps a few $e$-folds of thermal inflation tacked on later
to mop up any unwanted relics. 

Under the inflaton hypothesis this set-up is mandatory, but 
adopting instead the curvaton hypothesis things are not so clear.
Might it be that the curvaton field acquires its inhomogeneity during
an era of thermal inflation, at least on cosmological scales?
After a very few $e$-folds, thermal inflation certainly is 
of the needed almost-exponential
type, since the radiation density $\rho_\gamma$ falls like $T^{-4}\propto
a^{-4}$ leading to
\begin{equation}
\epsilon\simeq -\frac{\dot{H}}{H^2}=\frac{2\rho_\gamma}{\rho_\gamma+V_0}=
\frac{2}{1+\exp(4\Delta N_{\rm therm})}
\,.
\end{equation}
However one needs to 
ask whether  cosmological scales can  leave the horizon during
thermal inflation. 



\subsection{Ordinary and modular thermal inflation}

Two sorts of thermal inflation have been considered,
 depending on whether
$\phi$ is an ordinary field (`matter field' in the terminology of
string theory) or a modulus. 
For ordinary thermal inflation \cite{thermal2}
the temperature-dependent  effective potential is
\be
V(\phi,T)  =  V_0 +
(g T^2- \frac12 m^2)\phi^2+\frac{1}{d}\lambda\frac{\phi^d}{\mpl ^{d-4}}
\label{thermpot}
\ee
where $d>4$ is an integer, $g\sim 1$ is the coupling of 
$\phi$ with the particles of the thermal bath and $\lambda \gsim 1$
is the coupling of the leading non-renormalizable term.\footnote
{In terms of the ultra-violet cutoff $\Lambda \lsim \mpl$ of the effective
field theory the expected value is $\lambda \sim (\mpl/\Lambda)^{d-4}
\gsim 1$.}
This
 expression is supposed
to be a good approximation when $\phi$ is less than its VEV $M_0$.

We consider first the zero-temperature potential. Setting its derivative
equal to zero gives  the VEV
\begin{equation}
M_0=\Big(\frac{1}{\lambda}\Big)^{\frac{1}{d-2}}
\Big(\frac{m}{\mpl }\Big)^{\frac{2}{d-2}}\mpl  \ll\mpl
\,,
\label{M0}
\end{equation}
and evaluating the second derivative gives the
 mass in the vacuum as
\begin{equation}
m_\phi^2 = (d-2) m^2\sim m^2
\label{mphiapp}
\,.
\end{equation}
Demanding that \mbox{$V(M_0)=0$} gives 
\begin{equation}
V_0
=\frac{d-2}{2d}m^2M_0^2\sim m^2 M_0^2
\label{vapp}
\,.
\end{equation}
This gives
\begin{equation}
|\eta| = \frac{2d}{d-2} \(\frac{\mpl}{M_0} \)^2 
\sim  \mpl^2/M_0^2\gg 1
\label{etaapp}
\,.
\end{equation}
where $\eta=-m^2\mpl^2/V_0$.

For modular thermal inflation \cite{hs}, 
$\phi$ is supposed to be a 
modulus, with the zero-temperature potential
considered in Sections \ref{modular} and \ref{sfasrol}.
 In this case the last term of \eq{thermpot}
is replaced by some unknown function, but on the basis of string theory
examples the order of magnitude estimates of 
\eqss{mphiapp}{vapp}{etaapp} are assumed to be 
valid. It follows that for modular inflation, $|\eta|\sim 1$.

Now consider the evolution of $\phi$.
At $T$ bigger than $T_c\equiv (m/\sqrt{2 g})$, the effective mass-squared
$m^2(T) \equiv 2g T^2-m^2$ holds
$\phi$ close to the origin, and the energy density is
\be
\rho = \frac{\pi^2}{30} g_* T^4 + V_0
\,,
\ee
where
\mbox{$g_*$} is the effective number of relativistic degrees of 
freedom.
Thermal inflation starts  
when the second term starts to dominate at temperature
\begin{equation}
T\sub{start} = \(\frac{30}{\pi^2g_*}\)^{1/4}
V_0^{1/4} \sim V_0^{1/4}
\label{T0}
\end{equation}
Thermal inflation ends when $T=T_c$.
To estimate the number $N\sub{therm}$ of $e$-folds of thermal
inflation, we can take 
$g_*$ to be constant, giving $T\propto 1/a$ and
\be
N\sub{therm}\simeq \ln(V_0^{1/4}/m) =  
\ln(\mpl/V_0^{1/4}) - \frac12\ln|\eta|
\label{ntherm}
\,.
\ee

During thermal inflation, the typical field value is $\phi\sim T$.
The  value at the end of thermal inflation is therefore
$\phi\sim m$, and
before this value changes much the
potential has practically its zero temperature form. Then $\phi$
rolls away from the origin, reaching its VEV after
a number $N\sub{roll}$ of Hubble times given by
\eq{nroll}. 
For ordinary thermal inflation, $|\eta|$ is exponentially large and 
$N\sub{roll}\ll 1$. In this case there is no more inflation
after thermal inflation ends. In contrast, for modular inflation
$|\eta|$ is of order unity and one may have \mbox{$N\sub{roll}\gg 1$}.
In this case there are $N\sub{roll}$ $e$-folds of inflation after 
thermal inflation ends.

\subsection{\boldmath Thermal inflation with $m\ll H$}

To complete this discussion of the dynamics of thermal inflation, we need
to consider the case $|\eta|\ll 1$ or equivalently
$m \ll H$. In order to avoid $M_0 \gg 
\mpl$, the form \eq{thermpot} must in this case be modified so as to steepen
the potential, either in the $\phi$ direction or in the direction of some
other field as in inverted hybrid inflation \cite{Lyth:1996kt}. 

With $m \ll H$ there is a regime of temperature $m \ll T \ll H$.\footnote
{By analogy with the case of particles in equilibrium, 
the field $\phi$ will presumably fall out of thermal equilibrium 
when the temperature falls below $H$. But also by analogy with that case,
one can expect that
the form of the effective potential will continue to be the same as if there
were equilibrium. One can verify this explicitly for the case of 
thermal equilibrium with a scalar field $\chi$ through a coupling 
$g\phi^2\chi^2$, where the 
thermal average $\chi^2\sim T^2$ corresponds to the
contributions of plane waves representing  relativistic particles in thermal
equilibrium.} Before $T$ enters this regime, the 
effective mass-squared $\sim T^2$ is big enough to hold $\phi$ at the typical
value $\phi\sim T$ mentioned earlier. Afterwards though, the effective 
mass-squared falls below $H^2$, and 
$\phi$ begins a random
walk under the influence of the quantum fluctuation, moving a distance
$\pm H/2\pi$ during each Hubble time. This
continues until the potential becomes steep
enough that the random walk is slower than the classical roll given 
by \eq{sr}.\footnote
{If the modification drives the motion in the direction of another 
field it is the classical motion in that direction that is relevant.
We are discounting  the possibility
that the modification of the potential becomes significant before
$T\sim H$ since this would lead to a completely different type of
model.}
The random-walk era is therefore an era of `eternal' inflation
(so-called because its duration, in a given region, can be indefinitely long).
On scales leaving the horizon during eternal inflation, the curvature
perturbation is  of order 1, and when these  scales start to  enter
 the horizon
around half of the energy density of the Universe collapses
to black holes.

When eternal inflation has been considered previously, it has been 
supposed to occur before cosmological scales leave the horizon.
Then the scales on which the curvature perturbation is of order 1 
are outside the horizon at the present epoch and can be ignored
(except for the Grishchuk-Zeldovich effect which is excluded by
observation \cite{book}). In our case, we want cosmological scales to 
leave the horizon during thermal inflation and the eternal inflation
would generate a curvature perturbation of order 1 on sub-cosmological
scales. The black hole formation would, therefore be a disaster, generating
an irrevocably matter-dominated early Universe. (We discount the possibility
that the black holes evaporate, which would require them to form on a
 scale which is implausibly small in the present context.)
We conclude that thermal inflation with $m\ll H$ is unviable if cosmological
scales are required to exit the horizon during this inflation.\footnote{Note
that, when $m\gsim H$, there is also a brief period in which $m^2(T)<H^2$
 but it lasts less than a Hubble time so that there is no black hole 
formation.}

\subsection{Generating the curvature perturbation}

We now ask if cosmological scales can leave the horizon during thermal 
inflation, so that the curvaton may acquire its perturbation then. As we 
have discussed in Sec.~\ref{sfasrol} the biggest cosmological scale is given by
Eq.~(\ref{nbig}). We now impose the requirement that cosmological scales 
leave the horizon during thermal inflation.

Consider first ordinary thermal inflation,
where no more inflation takes place
after thermal inflation. 
In this case \eqs{ntherm}{nbig} show that,
regardless of $V_0$, the requirement  is $\ln(\mpl /m )\gsim 72-\Delta$.
If one assumes prompt reheating and $N_0=0$, then $\Delta=0$ and the above 
bound corresponds to
\bea
m &\lsim &10^{-4}\eV \label{mbound}\\
V_0^{1/4} &\lsim &\TeV \label{vbound}
\eea
It is easy to verify that practically the same bound is obtained even
if inefficient reheating is allowed, due to the the combined effect of the
requirement $|\eta|\gsim 1$ and 
the nucleosynthesis bound $T\sub{reh}\gsim 10\MeV$.
Such a small mass is completely unviable because it would prevent
reheating after thermal inflation. The only effective decay channel would be
to the photon with rate $\Gamma\sim m^3/\mpl^2$. Thus, since 
\mbox{$T_{\rm reh}\sim\sqrt{\Gamma\mpl}$}, we find\footnote{Even with 
$\Delta\neq 0$ one obtains $T_{\rm reh}\sim 10^{-19+9(\Delta/15)}{\rm eV}<
10^{-10}$eV.}

\begin{equation}
T_{\rm reh}\sim\sqrt{\frac{m^3}{\mpl}}\sim 10^{-19}{\rm eV}
\end{equation}
Therefore,
we conclude that {\em cosmological
scales cannot leave the horizon during ordinary thermal inflation}.

Now consider modular inflation, where  some number
$N\sub{roll}$ of inflationary e-folds take place after thermal inflation, 
given by \eq{nroll}. Since it takes about $\Delta N_{\rm cosm}\simeq 17$
$e$-folds for cosmological scales to leave the horizon, and we want
this to happen during thermal inflation, we require
\be
N_{\rm big}-N_{\rm therm}<N_{\rm roll}<N_{\rm big}-\Delta N_{\rm cosm}
\ee
or
\be
\frac{72-\Delta}{2+\frac2F} < \ln\left(\frac{\mpl}{V_0^{1/4}}\right)
< \frac{55-\Delta}{1+\frac2F}
\ee

There is a solution only for
\be
F> \frac{34}{38-\Delta}
\label{fbound}
\ee
(For smaller values of $F$,  cosmological scales
start to leave the horizon only after thermal inflation is over.)
With $\Delta=0$ this gives $|\eta| > 1.16$, and if $\Delta$ saturates
\eq{Dbound} it gives $|\eta| > 2.21$.

In the limit where the 
bound \eq{fbound} on $F$ is  saturated, we find independently of $\Delta$,
\be
\ln\left(\frac{\mpl}{V_0^{1/4}}\right) \simeq 17
\ee
corresponding
to
\bea
V_0^{1/4} &\simeq &1\times 10^{11}\GeV \label{v11} \\
m &\simeq & 4-6\;\TeV  \label{mtev}
\eea
Increasing $F$  decreases $V_0^{1/4}$. The limit
$F\to\infty$ corresponds to $V_0^{1/4}\sim \TeV$, in agreement with
\eq{vbound}.

The values represented by \eqs{v11}{mtev}, 
are about
 the ones expected for a modulus in the case of gravity-mediated
supersymmetry breaking. Our conclusion therefore is that cosmological
scales can leave the horizon during modular thermal inflation.

For any kind of modular inflation (slow-roll, fast-roll or
thermal) the subsequent oscillation of the modulus around its is
VEV must decay well before nucleosynthesis. The  usual assumption
is that the modulus mass is of order $m_{3/2}$ with the modulus 
decaying with  only gravitational strength, which requires
$m_{3/2} \gsim 10\TeV$ corresponding to 
anomaly-mediated SUSY breaking \cite{anmed}. 
This is mildly inconsistent with \eq{mtev}. However, it is conceivable 
\cite{hs} that the modulus involved with modular inflation  correspond to a
flat direction of the MSSM, in which case the decay would be prompt. 
Of course, in the case of thermal inflation this would require
that the interactions holding the modulus at its maximum
be different from the Standard Model interactions.

\subsection{The vacuum assumption during thermal inflation}

In the previous subsection, we have implicitly assumed that 
at the beginning of inflation, the curvaton field is in the vacuum
on scales within the horizon. This is because we assumed that
cosmological scales could start to leave the horizon as soon as 
thermal inflation starts, the vacuum assumption then 
being necessary to obtain the  flat spectrum for the 
curvaton field in the usual way.\footnote
{We shall use the term `vacuum state' only
for scales well inside the horizon, where it is unambiguously defined.
We do not consider the possibility that a flat spectrum could be
obtained starting from some  non-vacuum state, since no such state
has ever been exhibited.} We now ask to what extent this assumption is
justified.

There is a model-independent upper limit on the number 
$N\sub{bef}$ of $e$-folds
which elapse before scales leaving the horizon are the vacuum,
from the same consideration that has been invoked under the inflaton
hypothesis
 \cite{Liddle:1993fq}. At the beginning of
inflation, all scalar fields with mass $m\ll H$ 
must be in the vacuum on scales $k/a>V_0^{1/4}$ because they
would otherwise dominate the energy density and (having positive
pressure) spoil inflation. The upper limit on $N\sub{bef}$ is therefore
the number of $e$-folds which elapse before such scales start to leave
the horizon,
\be
N\sub{bef} < \ln\Big(V_0^{1/4}/H\Big)\sim \ln\Big(\mpl/V_0^{1/4}\Big)
\ee
But this uses up all of the $e$-folds of thermal inflation,
which blocks the desired objective that cosmological scales leave the
horizon during thermal inflation. We need fewer $e$-folds! On the other
hand, the following argument  shows that $N\sub{bef}$
 cannot be zero.

Suppose first that 
 the radiation domination era
which precedes thermal inflation is itself preceded by an inflationary
era. That, presumably, must be the case if the radiation dominated 
Universe is flat and  homogeneous to high accuracy. The reason that
$N\sub{bef}$  cannot then be zero is that during radiation domination
scales are entering the horizon which {\em left} the horizon during
the previous inflation. Such scales cannot be in the vacuum because their
vacuum fluctuation has been converted to a classical perturbation. 
However, the fate of such perturbations after reentry is not clear.
It is plausible that when the perturbations become again causally connected
their dynamics will result in suppression of their power at small scales,
even though they are weakly coupled.

 


Another possibility is that there is no substantial earlier inflation and
the Universe is only roughly flat, homogeneous and isotropic during the 
early radiation dominated era. Still, even in that case $N\sub{bef}$ 
has to be non-zero. This is because we must wait a few $e$-folds before the
 the observable Universe leaves the 
horizon so that it will have the observed extreme flatness, 
and homogeneity.

{}From this discussion we learn that some number of $e$-folds must
elapse before cosmological scales leave the horizon. On the other hand
we have not found any definite lower limit on this number, so that the
estimates that we earlier made by setting it equal to zero may still
be valid as an approximation.

 \section{Conclusion}
\label{sconc}
The first viable inflation models, generally termed
`new inflation' models \cite{Linde:1981mu},   involved a single field, the inflaton,
which was supposed to perform three tasks; support inflation, end inflation
and generate the curvature perturbation. Such a paradigm is 
economical in terms of the number of fields involved, 
but it makes model-building 
a quite difficult task and arguably involves unacceptable fine-tuning.

The hybrid inflation paradigm \cite{Linde:1991km} 
delegated the task of supporting inflation
(i.e., of generating the required potential) to `waterfall'  field
different from the inflaton.
 Hybrid inflation models are much easier
to construct  and can be relatively free of fine-tuning, but they are 
still quite constrained.

In both of these cases, the hypothesis regarding the origin of the 
curvature perturbation is the same; it comes from the perturbation
of the inflaton field.
The curvaton hypothesis is that 
the curvature perturbation comes from the perturbation of
 a different field, the curvaton. According
to this hypothesis, inflation itself may be of the `new' or `hybrid'
variety. (In the hybrid  case though, the curvaton cannot be identified
with the waterfall  field since the latter has mass much bigger than
$H$.) Alternatively, inflation  may be of a type 
not involving  any rolling field, such as thermal inflation.
In this paper, we have revisited most of the inflation models that have
been proposed, asking to what extent they become more attractive when
they are liberated by the curvaton hypothesis.

For slow-roll inflation, the results are summarized in Table 1. 
An interesting finding is that a modulus becomes a
more attractive candidate.
Turning to thermal inflation, we found again an attractive  model
involving a modulus.
When thermal inflation ends, some more
$e$-folds of inflation occur while the modulus rolls to the vacuum.
The model is perhaps more attractive than modular inflation  without
thermal inflation, because the initial value of the modulus is explained
and because there need not be so many $e$-folds of thermal inflation. 

 \subsection*{Acknowledgments}
DHL acknowledges  conversations with Andrei Lukas about TeV-scale 
inflation, and correspondence with Joe Polchinski about the
string moduli. This work was supported by the EU Fifth Framework network 
"Supersymmetry and the Early Universe" HPRN-CT-2000-00152 and also, 
in part, by the National Science Foundation under Grant No. PHY99-07949.
We are grateful to KITP, UCSB for hospitality.

 \newcommand\pl[3]{Phys.\ Lett.\ {\bf #1}  (#3) #2}
 \newcommand\np[3]{Nucl.\ Phys.\ {\bf #1}  (#3) #2}
 \newcommand\pr[3]{Phys.\ Rep.\ {\bf #1}  (#3) #2}
 \newcommand\prl[3]{Phys.\ Rev.\ Lett.\ {\bf #1}  (#3)  #2}
 \newcommand\prd[3]{Phys.\ Rev.\ D{\bf #1}  (#3) #2}
 \newcommand\ptp[3]{Prog.\ Theor.\ Phys.\ {\bf #1}  (#3)  #2 }
 \newcommand\rpp[3]{Rep.\ on Prog.\ in Phys.\ {\bf #1} (#3) #2}
 \newcommand\jhep[2]{JHEP #1 (#2)}
 \newcommand\grg[3]{Gen.\ Rel.\ Grav.\ {\bf #1}  (#3) #2}
 \newcommand\mnras[3]{MNRAS {\bf #1}   (#3) #2}
 \newcommand\apjl[3]{Astrophys.\ J.\ Lett.\ {\bf #1}  (#3) #2}


\begin{thebibliography}{99}

\bibitem{wmapparam}
D.~N.~Spergel {\it et al.},
Astrophys.\ J.\ Suppl.\  {\bf 148}, 175 (2003).

\bibitem{wmapsdss}
M.~Tegmark {\it et al.}  [SDSS Collaboration],
arXiv:astro-ph/0310723.

\bibitem{book} D. H. Lyth \& A. R. Liddle, {\sl Cosmological Inflation
 and Large-Scale Structure}, Cambridge University Press (2000).
 
\bibitem{treview}
D.~H.~Lyth and A.~Riotto,
Phys.\ Rept.\  {\bf 314}, 1 (1999).

\bibitem{bg} 
P.~Binetruy and M.~K.~Gaillard,
Phys.\ Rev.\ D {\bf 34}, 3069 (1986).

\bibitem{thermal1}
G.~Lazarides, C.~Panagiotakopoulos and Q.~Shafi,
Phys.\ Rev.\ Lett.\  {\bf 56}, 557 (1986).

\bibitem{thermal2} 
D.~H.~Lyth and E.~D.~Stewart,
Phys.\ Rev.\ Lett.\  {\bf 75}, 201 (1995);
%
D.~H.~Lyth and E.~D.~Stewart,
Phys.\ Rev.\ D {\bf 53}, 1784 (1996);
%
T.~Barreiro, E.~J.~Copeland, D.~H.~Lyth and T.~Prokopec,
Phys.\ Rev.\ D {\bf 54}, 1379 (1996).

\bibitem{thermalrest}
E.~D.~Stewart, M.~Kawasaki and T.~Yanagida,
Phys.\ Rev.\ D {\bf 54} (1996) 6032;
%
T.~Asaka and M.~Kawasaki,
Phys.\ Rev.\ D {\bf 60} (1999) 123509;
%
R.~Jeannerot,
Phys.\ Rev.\ D {\bf 59} (1999) 083501;
%
T.~Asaka, M.~Kawasaki and T.~Yanagida,
Phys.\ Rev.\ D {\bf 60} (1999) 103518.

\bibitem{hs}
L.~Hui and E.~D.~Stewart,
Phys.\ Rev.\ D {\bf 60} (1999) 023518.

\bibitem{Lyth:2001nq}
D.~H.~Lyth and D.~Wands,
Phys.\ Lett.\ B {\bf 524}, 5 (2002).

\bibitem{sylvia} 
S.~Mollerach,
Phys.\ Rev.\ D {\bf 42}, 313 (1990).

\bibitem{lm}
A.~D.~Linde and V.~Mukhanov,
Phys.\ Rev.\ D {\bf 56}, 535 (1997).

\bibitem{mt1} 
T.~Moroi and T.~Takahashi,
Phys.\ Lett.\ B {\bf 522}, 215 (2001).

\bibitem{andrew} 
N.~Bartolo and A.~R.~Liddle,
Phys.\ Rev.\ D {\bf 65}, 121301 (2002).

\bibitem{mt2} 
T.~Moroi and T.~Takahashi,
arXiv:hep-ph/0206026.

\bibitem{fy}
M.~Fujii and T.~Yanagida,
Phys.\ Rev.\ D {\bf 66} 123515,  (2002).

\bibitem{luw}
D.~H.~Lyth, C.~Ungarelli and D.~Wands,
arXiv:astro-ph/0208055.

\bibitem{sloth}
M.~S.~Sloth,
Nucl.\ Phys.\ B {\bf 656}, 239 (2003).

\bibitem{hmy}
A.~Hebecker, J.~March-Russell and T.~Yanagida,
arXiv:hep-ph/0208249.

\bibitem{hofmann} 
R.~Hofmann,
arXiv:hep-ph/0208267.

\bibitem{mormur}
T.~Moroi and H.~Murayama,
arXiv:hep-ph/0211019.

\bibitem{ekm} 
K.~Enqvist, S.~Kasuya and A.~Mazumdar,
arXiv:hep-ph/0211147.

\bibitem{mwu} 
K.~A.~Malik, D.~Wands and C.~Ungarelli,
arXiv:astro-ph/0211602.

\bibitem{postma} 
M.~Postma,
arXiv:hep-ph/0212005.

\bibitem{fl}
B.~Feng and M.~Li, arXiv:hep-ph/0212213.

\bibitem{gl}
C.~Gordon and A.~Lewis,
Phys.\ Rev.\ D {\bf 67},   123513  (2003).

\bibitem{kostasquint}
K.~Dimopoulos,
arXiv:astro-ph/0212264.

\bibitem{giov}
M.~Giovannini,
Phys.\ Rev.\ D {\bf 67} (2003) 123512.

\bibitem{lu}
A.~R.~Liddle and L.~A.~Urena-Lopez,
arXiv:astro-ph/0302054.

\bibitem{dllr1}
K.~Dimopoulos, G.~Lazarides, D.~Lyth and R.~Ruiz de Austri,
arXiv:hep-ph/0303154.

\bibitem{ejkm}
K.~Enqvist, A.~Jokinen, S.~Kasuya and A.~Mazumdar,
arXiv:hep-ph/0303165.

\bibitem{03dl2}
D.~H.~Lyth and D.~Wands,
arXiv:astro-ph/0306500.

\bibitem{dllr2}
K.~Dimopoulos, G.~Lazarides, D.~Lyth and R.~Ruiz~de Austri,
arXiv:hep-ph/0308015.

\bibitem{dlnr}
K.~Dimopoulos, D.~H.~Lyth, A.~Notari and A.~Riotto,
JHEP {\bf 0307} (2003) 053.

\bibitem{pm}
M.~Postma and A.~Mazumdar,
arXiv:hep-ph/0304246.

\bibitem{john2}
J.~McDonald,
arXiv:hep-ph/0308048.

\bibitem{kkt}
S.~Kasuya, M.~Kawasaki and F.~Takahashi,
arXiv:hep-ph/0305134.

\bibitem{ekm03}
M.~Endo, M.~Kawasaki and T.~Moroi,
arXiv:hep-ph/0304126.

\bibitem{hktt}
K.~Hamaguchi, M.~Kawasaki, T.~Moroi and F.~Takahashi,
arXiv:hep-ph/0308174.

\bibitem{john3}
J.~McDonald,
arXiv:hep-ph/0308295.

\bibitem{bmr}
N.~Bartolo, S.~Matarrese and A.~Riotto,
arXiv:hep-ph/0309033.

\bibitem{bmr2}
N.~Bartolo, S.~Matarrese and A.~Riotto,
arXiv:astro-ph/0309692.

\bibitem{giovannini03}
M.~Giovannini,
arXiv:hep-ph/0310024.

\bibitem{john4}
J.~McDonald,
arXiv:hep-ph/0310126.

\bibitem{mazumdar}
A.~Mazumdar,
arXiv:hep-th/0310162.

\bibitem{ad}
M.~Axenides and K.~Dimopoulos,
arXiv:hep-ph/0310194.

\bibitem{armen}
C.~Armendariz-Picon,
arXiv:astro-ph/0310512.

\bibitem{cdl}
E.\ J.\ Chun, K.\ Dimopoulos \& D.\ H.\ Lyth, in preparation.

\bibitem{decay}
G.~Dvali, A.~Gruzinov and M.~Zaldarriaga,
arXiv:astro-ph/0303591;
L.~Kofman,
arXiv:astro-ph/0303614;
K.~Enqvist, A.~Mazumdar and M.~Postma,
inflaton coupling,''
Phys.\ Rev.\ D {\bf 67}, 121303 (2003)
[arXiv:astro-ph/0304187];
G.~Dvali, A.~Gruzinov and M.~Zaldarriaga,
and Mass Domination,''
arXiv:astro-ph/0305548;
S.~Tsujikawa,
arXiv:astro-ph/0305569;
S.~Matarrese and A.~Riotto,
the inflaton decay rate,''
arXiv:astro-ph/0306416;
A.~Mazumdar and M.~Postma,
arXiv:astro-ph/0306509.

\bibitem{steve2} 
M.~Bastero-Gil, V.~Di Clemente and S.~F.~King,
arXiv:hep-ph/0211011;
M.~Bastero-Gil, V.~Di Clemente and S.~F.~King,
arXiv:hep-ph/0211012;
M.~Bastero-Gil, V.~Di Clemente and S.~F.~King, in preparation.

\bibitem{p00n1} 
D.~H.~Lyth and L.~Covi,
Phys.\ Rev.\ D {\bf 62}, 103504 (2000).

\bibitem{gw02} 
L.~Knox and Y.~S.~Song,
Phys.\ Rev.\ Lett.\  {\bf 89}, 011303 (2002);
%
M.~Kesden, A.~Cooray and M.~Kamionkowski,
Phys.\ Rev.\ Lett.\  {\bf 89}, 011304 (2002).

\bibitem{wein} 
S.~Weinberg,
{\sl The Quantum Theory Of Fields.  Vol. 3: Supersymmetry},
Cambridge University Press (2000).


\bibitem{pol} 
J.~Polchinski,
{\sl String Theory. Vol. 2: Superstring Theory And Beyond},
Cambridge University Press (1998).

\bibitem{cllsw} 
E.~J.~Copeland, A.~R.~Liddle, D.~H.~Lyth, E.~D.~Stewart and D.~Wands,
Phys.\ Rev.\ D {\bf 49}, 6410 (1994).

\bibitem{glm} 
M.~K.~Gaillard, D.~H.~Lyth and H.~Murayama,
Phys.\ Rev.\ D {\bf 58}, 123505 (1998).

\bibitem{km} 
C.~F.~Kolda and J.~March-Russell,
Phys.\ Rev.\ D {\bf 60}, 023504 (1999).

\bibitem{modular1}
T.~Banks, M.~Berkooz, S.~H.~Shenker, G.~W.~Moore and P.~J.~Steinhardt,
Phys.\ Rev.\ D {\bf 52}, 3548 (1995).

\bibitem{modular2}
T.~Banks,
arXiv:hep-th/9906126;
%
R.~Brustein, S.~P.~De Alwis and E.~G.~Novak,
 space,''
arXiv:hep-th/0205042.

\bibitem{natural2}
F.~C.~Adams, J.~R.~Bond, K.~Freese, J.~A.~Frieman and A.~V.~Olinto,
Phys.\ Rev.\ D {\bf 47}, 426 (1993).

\bibitem{natural1}
K.~Freese, J.~A.~Frieman and A.~V.~Olinto,
Phys.\ Rev.\ Lett.\  {\bf 65}, 3233 (1990).

\bibitem{topological}
A.~Vilenkin,
Phys.\ Rev.\ Lett.\  {\bf 72}, 3137 (1994);
%
A.~D.~Linde and D.~A.~Linde,
Phys.\ Rev.\ D {\bf 50}, 2456 (1994).

\bibitem{anmed}
L.~Randall and R.~Sundrum,
Nucl.\ Phys.\ B {\bf 557}, 79 (1999);
%
T.~Moroi and L.~Randall,
Nucl.\ Phys.\ B {\bf 570}, 455 (2000).

\bibitem{Cohn:2000hc}
J.~D.~Cohn and E.~D.~Stewart,
Phys.\ Lett.\ B {\bf 475}, 231 (2000).

\bibitem{Stewart:2000pa}
E.~D.~Stewart and J.~D.~Cohn,
 gauge symmetries,''
Phys.\ Rev.\ D {\bf 63}, 083519 (2001).

\bibitem{Dine:1997kf}
M.~Dine and A.~Riotto,
Phys.\ Rev.\ Lett.\  {\bf 79}, 2632 (1997).

\bibitem{Dine:1995ag}
M.~Dine, A.~E.~Nelson, Y.~Nir and Y.~Shirman,
Phys.\ Rev.\ D {\bf 53}, 2658 (1996).

\bibitem{Lyth:1999ty}
D.~H.~Lyth,
Phys.\ Lett.\ B {\bf 466}, 85 (1999).

\bibitem{Linde:1991km}
A.~D.~Linde,
Phys.\ Lett.\ B {\bf 259}, 38 (1991).

\bibitem{dss}
G.~R.~Dvali, Q.~Shafi and R.~Schaefer,
Phys.\ Rev.\ Lett.\  {\bf 73}, 1886 (1994).

\bibitem{ewan} 
E.~D.~Stewart,
Phys.\ Rev.\ D {\bf 51}, 6847 (1995).

\bibitem{bd}
P.~Binetruy and G.~R.~Dvali,
Phys.\ Lett.\ B {\bf 388}, 241 (1996).

\bibitem{halyo}
E.~Halyo,
Phys.\ Lett.\ B {\bf 387}, 43 (1996).

\bibitem{running} 
E.~D.~Stewart,
Phys.\ Lett.\ B {\bf 391}, 34 (1997);
%
E.~D.~Stewart,
Phys.\ Rev.\ D {\bf 56}, 2019 (1997).

\bibitem{Covi:1998mb}
L.~Covi, D.~H.~Lyth and L.~Roszkowski,
Phys.\ Rev.\ D {\bf 60}, 023509 (1999);
%
L.~Covi and D.~H.~Lyth,
Phys.\ Rev.\ D {\bf 59}, 063515 (1999).

\bibitem{p00n2}
L.~Covi and D.~H.~Lyth,
  Mon. Not. Roy. Astr. Soc. {\bf 326}, 877 (2001).

\bibitem{p02laura}
L.~Covi, D.~H.~Lyth and A.~Melchiorri,
arXiv:hep-ph/0210395.

\bibitem{clmprep}
L.~Covi, D.~H.~Lyth and A.~Melchiorri, in preparation.

\bibitem{clprep}
L.~Covi and  D.~H.~Lyth, in preparation.

\bibitem{Bastero-Gil:1997vn}
M.~Bastero-Gil and S.~F.~King,
Phys.\ Lett.\ B {\bf 423}, 27 (1998).

\bibitem{Dvali:1998pa}
G.~R.~Dvali and S.~H.~Tye,
Phys.\ Lett.\ B {\bf 450}, 72 (1999).

\bibitem{Alexander:2001ks}
S.~H.~Alexander,
Phys.\ Rev.\ D {\bf 65}, 023507 (2002).

\bibitem{quevedo} 
C.~P.~Burgess, M.~Majumdar, D.~Nolte, F.~Quevedo, G.~Rajesh and R.~J.~Zhang,
JHEP {\bf 0107}, 047 (2001);
%
S.~Alexander, Y.~Ling and L.~Smolin,
 cosmologies,''
Phys.\ Rev.\ D {\bf 65}, 083503 (2002).

\bibitem{Dvali:2001fw}
G.~R.~Dvali, Q.~Shafi and S.~Solganik,
arXiv:hep-th/0105203.

\bibitem{Shiu:2001sy}
G.~Shiu and S.~H.~Tye,
Phys.\ Lett.\ B {\bf 516}, 421 (2001).

\bibitem{La:za}
D.~La and P.~J.~Steinhardt,
Phys.\ Rev.\ Lett.\  {\bf 62}, 376 (1989)
[Erratum-ibid.\  {\bf 62}, 1066 (1989)].

\bibitem{Starobinsky:te}
A.~A.~Starobinsky,
Phys.\ Lett.\ B {\bf 91}, 99 (1980).

\bibitem{Hawking:2000bb}
S.~W.~Hawking, T.~Hertog and H.~S.~Reall,
Phys.\ Rev.\ D {\bf 63}, 083504 (2001).

\bibitem{Linde:2001ae}
A.~Linde,
JHEP {\bf 0111}, 052 (2001).

\bibitem{Stewart:1993bc}
E.~D.~Stewart and D.~H.~Lyth,
Phys.\ Lett.\ B {\bf 302}, 171 (1993).

\bibitem{Randall:1995dj}
L.~Randall, M.~Soljacic and A.~H.~Guth,
Nucl.\ Phys.\ B {\bf 472}, 377 (1996).

\bibitem{Lyth:1996kt}
D.~H.~Lyth and E.~D.~Stewart,
Phys.\ Rev.\ D {\bf 54}, 7186 (1996).

\bibitem{Liddle:1993fq}
A.~R.~Liddle and D.~H.~Lyth,
Phys.\ Rept.\  {\bf 231}, 1 (1993).

\bibitem{Linde:1981mu}
A.~D.~Linde,
Phys.\ Lett.\ B {\bf 108}, 389 (1982).
%
A.~Albrecht and P.~J.~Steinhardt,
Phys.\ Rev.\ Lett.\  {\bf 48}, 1220 (1982).



 \end{thebibliography}
 \end{document}